\documentclass[12pt]{article}
\usepackage{amssymb,hyperref,graphicx}

\newcommand{\beq}{\begin{equation}}
\newcommand{\eeq}[1]{\label{#1}\end{equation}}
\newcommand{\be}{\begin{equation}}
\newcommand{\ee}{\end{equation}}
\newcommand{\bea}{\begin{eqnarray}}
\newcommand{\eea}{\end{eqnarray}}
\newcommand{\beas}{\begin{eqnarray*}}
\newcommand{\eeas}{\end{eqnarray*}}

\begin{document}
\begin{titlepage}


\medskip

\begin{center}

{\Large Back to the Future: Causality on a Moving Braneworld}

\vspace{12mm}

\renewcommand\thefootnote{\mbox{$\fnsymbol{footnote}$}}
Brian Greene${}^{1}$\footnote{brian.greene@columbia.edu},
Daniel Kabat${}^{2,3}$\footnote{daniel.kabat@lehman.cuny.edu},
Janna Levin${}^{4}$\footnote{janna@astro.columbia.edu},
Massimo Porrati${}^{5}$\footnote{massimo.porrati@nyu.edu}

\vspace{6mm}

${}^1${\small \sl Departments of Physics and Mathematics, Columbia University} \\
{\small \sl 538 West 120th Street, New York, NY 10027, USA}

\vspace{3mm}

${}^2${\small \sl Department of Physics and Astronomy} \\
{\small \sl Lehman College, City University of New York} \\
{\small \sl 250 Bedford Park Blvd.\ W, Bronx, NY 10468, USA}

\vspace{3mm}

${}^3${\small \sl Graduate School and University Center, City University of New York} \\
{\small \sl  365 Fifth Avenue, New York, NY 10016, USA}

\vspace{3mm}

${}^4${\small \sl Department of Physics and Astronomy} \\
{\small \sl Barnard College of Columbia University} \\
{\small \sl New York, NY 10027, USA}

\vspace{3mm}

${}^5${\small \sl Center for Cosmology and Particle Physics} \\
{\small \sl Department of Physics, New York University} \\
{\small \sl 726 Broadway, New York, NY 10003, USA}

\end{center}

\vspace{12mm}

\noindent
Brane observers executing appropriate motion through a partially compactified Lorentz invariant bulk spacetime, such as $M_4 \times S^1$, can send signals along the brane that are instantaneous or even travel backward in time. Nevertheless, causality in the braneworld remains intact. We establish these results, which follow from superluminal signal propagation reported in \cite{Greene:2022urm}, through classical analysis and then extend our reasoning by examining quantum mechanical microcausality. One implication is the capacity for real time communication across arbitrarily large distances.

\end{titlepage}

\setcounter{footnote}{0}
\renewcommand\thefootnote{\mbox{\arabic{footnote}}}

\hrule
\tableofcontents
\bigskip
\hrule

\addtolength{\parskip}{8pt}
\section{Introduction\label{sect:intro}}

Consider a 4D braneworld embedded in the bulk spacetime, $M_4 \times S^1$. If the brane is at rest on $S^1$, the brane worldvolume has an exact 4D Lorentz symmetry. But if the brane is moving, the Lorentz symmetry is broken globally by the compactification.  As shown
in \cite{Greene:2022urm} this allows superluminal signal propagation on the brane. Specifically, if the speed of the brane is $\beta$ (relative to the preferred rest frame on $S^1$, see section \ref{sect:frames}), \cite{Greene:2022urm} found that the effective speed of signal propagation on the brane is $\gamma = 1 / \sqrt{1 - \beta^2} \ge 1$.

One particularly useful way of understanding this result is to consider the covering space of  $M_4 \times S^1$, namely $M_5$, with coordinates $(t, x, {\bf y}, z)$.  Compactification amounts to periodically identifying the $z$ coordinate:
\be
z \sim z + 2 \pi R n \qquad n \in {\mathbb Z}
\ee
Then, consider a bulk light signal propagating in the $+x$ direction with trajectory
\bea
\label{trajectory}
&& x = t \\
\nonumber
&& {\bf y} = z = 0
\eea
Consider a brane that is extended in the $x$ and ${\bf y}$ directions and moving in the $z$ direction with velocity $\beta$.  In the covering space this corresponds to a series of branes located at $z = \beta t + 2 \pi R n$. The periodic identification
results in the light signal ``hopping"  from the original brane, located at $z = \beta t$, to the next image brane, located at $z = \beta t - 2 \pi R$.  From the brane point of view this is akin to the signal traveling from brane to image brane by traversing a connecting wormhole. More specifically, the signal leaves the brane at $t = x = 0$ and arrives at the image at $t = 2 \pi R / \beta$, $x = 2 \pi R / \beta$.  But clocks on a moving brane run slow.  Due to time dilation in the reference frame of the brane the signal reappears after a time
\be
\Delta t' = \Delta t / \gamma = 2 \pi R / \gamma \beta
\ee
The position is unchanged on the brane,
\be
\Delta x' = \Delta x = 2 \pi R / \beta
\ee
so according to an observer on the brane $\Delta x'  = \gamma \Delta t'$. When this ``hopping'' is repeated many times the signal propagates with an effective speed $v = \gamma > 1$.

An equivalent way of seeing this effect is to pass to comoving coordinates for the brane.
\bea
\nonumber
&& t' = \gamma(t - \beta z) \\
\label{S'}
&& x' = x \\
\nonumber
&& {\bf y'} = \bf{y}  \\
\nonumber
&& z' = \gamma(z - \beta t)
\eea
In these coordinates the brane is at $z' = 0$ but the identification becomes
\be
\label{S'_identification}
\left(\begin{array}{c} t' \\ z' \end{array} \right) \sim \left(\begin{array}{c} t' \\ z' \end{array} \right) +
\left(\begin{array}{c} - \gamma \beta 2 \pi R \\ \gamma 2 \pi R \end{array}\right)
\ee
In the new coordinates the signal's trajectory (\ref{trajectory}) is
\bea
\nonumber
&& t' = \gamma t \\
&& x' = t = t' / \gamma \\
\nonumber
&&z' = - \gamma \beta t = - \beta t'
\eea
After a time $t' = \gamma 2 \pi R / \beta$ the signal lands on the image brane at $z' = - \gamma 2 \pi R$.  This is identified with the original brane using (\ref{S'_identification}). So the signal arrives on the brane with location
$z' = - \gamma 2 \pi R + \gamma 2 \pi R = 0$ at position $x' = x = 2 \pi R / \beta$ and time $t' = \gamma 2 \pi R / \beta - \gamma \beta 2 \pi R = 2 \pi R / \gamma \beta$.  So again the effective speed is $v = \gamma$.

As is well known, superluminal signals can provide an affront to causality. In a Lorentz-invariant theory, some inertial observers will find that a superluminal signal travels backward in time, raising the specter of closed timelike curves. Of course, superluminality alone does not necessarily yield such causality-challenged trajectories and, moreover, in our setting Lorentz invariance on the brane is broken. But the analogy with wormholes alluded to above illuminates a potential concern. The roundtrip journey of a signal traversing two distinct wormholes, each propelling the signal back in time, can yield a closed timelike curve. Similarly, might there be trajectories of a signal ``hopping" from one image brane to another that can return to their source before they were emitted?

A quick argument suggests that this cannot happen. In the covering space, image points that are mutually identified all lie outside of each other's respective lightcones. This ensures that no image signal identified with that emitted from a given point can lie in that point's past lightcone. Nevertheless, the detailed dynamics of how superluminal signals on a moving brane preserve causality are not immediately apparent, and as we will see below, exploring them more fully provides significant and surprising insights.

\section{Preferred frames\label{sect:frames}}

From the standpoint of traditional Lorentz invariant theories on Minkowski space, an essential distinction that arises from compact spatial directions is that the associated global identifications pick out preferred frames of reference. For definiteness, focus on the simplest case, $M_4 \times S^1$. The preferred coordinate frame for the $S^1$ can be described mathematically as that frame for which the global identifications are purely spatial (equivalently, the preferred frame is the one in which there is a purely spatial Killing vector field whose integral curves form closed orbits). This frame also has a simple physical description: The preferred observers are those for whom two light signals emitted simultaneously and circumnavigating the $S^1$ in the clockwise and counterclockwise directions respectively, return simultaneously. Any observer moving relative to the preferred observers will find that such light signals do not arrive simultaneously and,  indeed, by determining the direction from which the first of the two light signal returns, the observer concludes that they are moving (relative to the preferred observer) in the opposite direction.

Let us now consider a 4D brane embedded in $M_4 \times S^1$, moving along the compact direction with speed $\beta$ relative to the preferred frame on $S^1$ that we just established. The claim is that the 4D brane itself has a preferred frame, one that is determined by the preferred frame on $S^1$. Namely, the preferred frame is the one in which the identifications (\ref{S'_identification}) are spatially perpendicular to the brane, and so when extended to the full spacetime take the form (here ${\bf x}' = (x',{\bf y}')$): 

\be
\label{S'_identificationextended}
\left(\begin{array}{c} t' \\ {\bf x'} \\ z' \end{array} \right) \sim \left(\begin{array}{c} t' \\ {\bf x'} \\ z' \end{array} \right) +
\left(\begin{array}{c} - \gamma \beta 2 \pi R \\ 0 \\ \gamma 2 \pi R \end{array}\right)
\ee
If we boost from this frame along the brane, setting
\bea
\nonumber
&&t' = \Gamma(t'' + B x'') \\
\label{S''}
&&x' = \Gamma(x'' + B t'') \\
\nonumber
&&{\bf y}' = {\bf y}'' \\
\nonumber
&&z' = z''
\eea
then the identification (\ref{S'_identification}) becomes
\be
\label{S''_identification}
\left(\begin{array}{c} t'' \\ x'' \\ z'' \end{array} \right) \sim \left(\begin{array}{c} t'' \\ x'' \\ z'' \end{array} \right) +
\left(\begin{array}{c} - \Gamma \gamma \beta 2 \pi R \\ \Gamma B \gamma \beta 2 \pi R \\ \gamma 2 \pi R \end{array}\right)
\ee
Note the simple but important point that in all but the preferred frame on the brane, the identification (\ref{S'_identification}) picks up a spatial component in the direction of the boost.  As we will see, this means for all but a preferred set of brane observers, bulk signals do not propagate isotropically on the brane.

\section{Bulk signals and signalling to the past\label{sect:bulk}}
In this section we consider a bulk signal sent out from $t = {\bf x} = z = 0$ and examine its future lightcone on the brane, establishing the possibility that according to brane observers the signal has been sent backwards
in time.  Unlike the introduction, where we considered a signal sent in the $+x$ direction, we're now imagining that the bulk signal is sent out isotropically.

Using the brane's preferred frame, $(t', {\bf x}', z')$, it was shown in
\cite{Greene:2022urm} that the future lightcone of the origin consists of an infinite series of images
labeled by $n \in {\mathbb Z}$, which in the covering space are given by
\be
\label{LightCones1}
\vert {\bf x}' \vert^2 + (z' - z_n')^2 = (t' + \beta z_n')^2
\ee
The $n^{th}$ image charge is located at $z_n' = \gamma 2 \pi R n$ and goes off at time $t_n' = - \beta z_n'$.
In coordinates $(t'', x'', {\bf y}'', z'')$ that are boosted along the brane the images become
\be
\label{images}
(x'' - \Gamma B \beta z_n')^2 + \vert {\bf y}'' \vert^2 + (z'' - z_n')^2 = (t'' + \Gamma \beta z_n')^2
\ee
The image with $n = 0$ is simply the lightcone of the origin.  For $n \not= 0$ the image is a spacelike hyperboloid, asymptotic to a lightcone emanating from
\bea
\nonumber
&&x'' = \Gamma B \beta z_n' \\
&&{\bf y}'' = 0 \\
\nonumber
&&z'' = z_n' \\
\nonumber
&&t'' = - \Gamma \beta z_n'
\eea
The individual images asymptotically expand at the speed of light on the brane.

\begin{figure}
\begin{center}
\includegraphics[height=6cm]{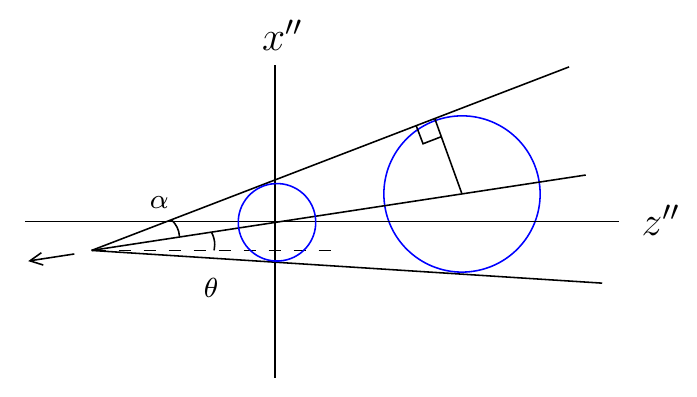}
\caption{Blue circles: light cones produced by image charges on a slice of constant $t''$.  The envelope forms a cone with opening angle $\alpha$ that moves in the indicated direction, along a line making an angle
$\theta$ with respect to the $z''$ axis.\label{fig:cone}}
\end{center}
\end{figure}

If we look at the images on a slice of fixed $t''$ we have a series of circles.  The centers of the circles lie along the line $x'' = \Gamma B \beta z''$.  This line
makes an angle $\theta$ with respect to the $z''$ axis where
\be
\label{tantheta}
\tan \theta = \Gamma B \beta
\ee
The radius of the $n^{th}$ circle is $t'' + \Gamma \beta z_n'$.  As shown in Fig.\ \ref{fig:cone} the envelope of the circles forms a cone.  The tip of the cone, where the radius shrinks to zero, is located at
\bea
&&z'' = - {t'' \over \Gamma \beta} \\
\nonumber
&&x'' = - B t''
\eea
The opening angle of the cone $\alpha$ is determined by
\be
\sin \alpha = {\hbox{\footnotesize radius of circle} \over \hbox{\footnotesize tip-to-center distance}}
\ee
and a bit of algebra shows that
\be
\label{sinalpha}
\sin^2 \alpha = {\beta^2 \over 1 - B^2 + \beta^2 B^2}
\ee
The opening angle is bounded, $0 < \alpha < \pi / 2$.  To see this note that the right hand side ranges from $0$ to a maximum value
\be
(1 - B^2)(1 - \beta^2) > 0 \quad \Rightarrow \quad {\beta^2 \over 1 - B^2 + \beta^2 B^2} < 1
\ee
The envelope forms angles $\theta \pm \alpha$ with respect to the $z''$ axis.
From (\ref{sinalpha}) we have $\tan \alpha = \Gamma \gamma \beta$ and comparing to (\ref{tantheta}) we see that $\alpha > \theta$.
In other words the lower part of the envelope always slopes downward while the upper part can have any angle between $0$ and $\pi$.

\begin{figure}
\begin{center}
\centerline{\includegraphics[height=7cm]{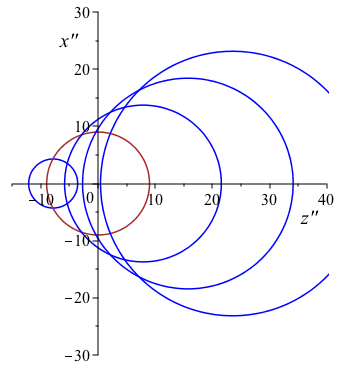} \hspace{6mm} \includegraphics[height=7cm]{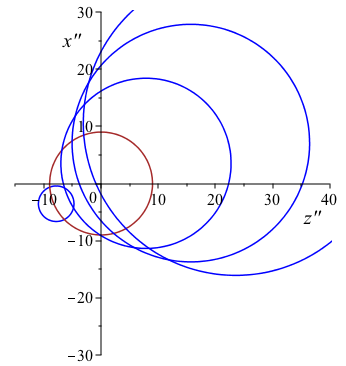}}
\caption{Light cones produced by image charges as seen on a slice of constant time, $t'' = 9$.  The brown circle is the light cone of the origin.  Shown in blue are the light cones for image charges $n = -1$ and $n =1,2,3$.  The left panel is in the
preferred brane frame, $B = 0$; note the symmetric superluminal spread of the signal in the $\pm x''$ direction.  The right panel has been boosted along the brane, $B = 0.6$.  Note the enhanced superluminal spread in the direction of
the boost.  In both cases $\beta = 0.6$ and $R = 1$.
\label{fig:flashes2}}
\end{center}
\end{figure}

At time $t''$ the envelope cuts the $x''$ axis at
\be
x'' = - B t'' + {t'' \over \Gamma \beta} \tan (\theta \pm \alpha)
\ee
This establishes the important fact that $x'' \ge t''$, which in turn shows that the envelope advances along the brane superluminally. Below, we will spell this out in more detail, but the snapshots in Figure \ref{fig:flashes2} provide a heuristic explanation of the essential physics. For positive values of $\beta$ and $B$, the sequence of image flashes located at ever larger values of $z''$ (that are all simultaneous in the preferred frame), leave their sources at ever earlier times. This provides an increasingly large temporal ``head start" for flashes at ever larger $z''$, which increases the maximal value of $x''$ the flashes can reach at a given $t''$. What's more, flashes at ever greater $z''$ (hence ever earlier $t''$) leave their sources at ever larger values of $x''$, (as is clear from  (\ref{S''_identification}) which, save the relativistic corrections, is just the spatial shift arising from the observer's motion in the positive $x''$ direction). This provides a spatial ``head start" for flashes at ever larger $z''$, which increases yet farther the maximal value of $x''$ the flashes can reach at a given time $t''$. Together, these two effects ensure an effective superluminal widening of the light cone sourced by image flashes at ever larger $z''$. This is evident in Figure \ref{fig:flashes2} where we see the successive intersections of the blue flashes with the brane reaching a greater distance from the origin than the brown flash.

\begin{figure}
\begin{center}
\centerline{\includegraphics[height=6cm]{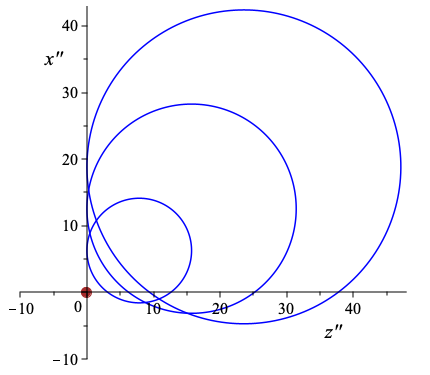}}
\caption{Illustrates the critical velocity.  The figure shows light cones produced by image charges at $t'' = 0$.  The light cone of the origin is just the point $z'' = x'' = 0$, indicated by a brown dot.  In blue are the light cones for image charges $n = 1,2,3$.  Note
the instantaneous signal propagation in the $+x''$ direction.  ($B = 0.8$, $\beta = 0.6$, $R = 1$)
\label{fig:flashes3}}
\end{center}
\end{figure}

In more detail, the intersection of the upper part of the envelope with the brane has a velocity
\bea
\nonumber
v & = & {\tan (\theta + \alpha) \over \Gamma \beta} - B \\[3pt]
\label{v}
& = & \gamma \, {1 + \Gamma^2 B^2 \beta^2 \over 1 - \Gamma^2 B \gamma \beta^2}
\eea
establishing that the light front along the brane in the direction we've boosted has a superluminal velocity. For $B = 0$, this reduces to the result $v = \gamma$ found in \cite{Greene:2022urm}. However, the case $B \ne 0$ offers new and particularly surprising possibilities. In particular, notice that for  $B_{\rm critical}$ defined by $\Gamma^2 B_{\rm critical} = 1/\gamma \beta^2$, the effective velocity of the light front diverges. That is, when we boost with speed $B_{\rm critical}$, we have instantaneous propagation along the brane, as illustrated in Figure \ref{fig:flashes3}.

\begin{figure}
\begin{center}
\centerline{\includegraphics[height=6cm]{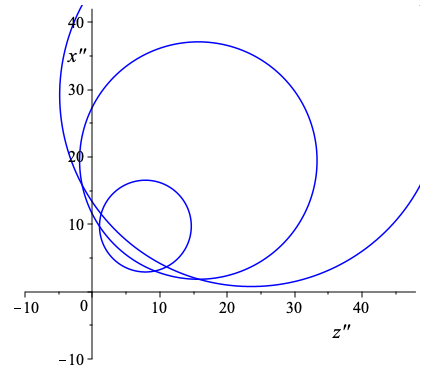}}
\caption{Illustrates a super-critical velocity.  The figure shows the $(z'',x'')$ plane at time $t'' = -4$.  The blue circles are the light cones produced by image charges $n = 1,2,3$.  The signal has been sent backwards in time and is propagating in the $-x''$ direction; it hasn't yet reached the origin where the signal will be emitted.  ($B = 0.9$, $\beta = 0.6$, $R = 1$)
\label{fig:flashes4}}
\end{center}
\end{figure}

For $B > B_{\rm critical}$ the signal propagation speed turns negative, which means the signal propagates
backwards in time as illustrated in Figure \ref{fig:flashes4}.  Perhaps the superluminal signal velocity renders this outcome inevitable, but we still find it remarkable that simply having an extra compact spatial direction in an otherwise Lorentz invariant theory yields a controlled classical setting in which observers can send signals to the past.

The extremal speed of propagation of the signal obtained in the limit of many ``hoppings'' can be obtained also by
considering the propagation of a signal with trajectory $x=t\cos\vartheta $, $z=t\sin\vartheta $, $\bf{y}=0$. Such a signal reaches
the first image brane, located at $z = \beta t - 2 \pi R$, at
\be
t = {2 \pi R \over \beta - \sin \vartheta} \qquad x = {2 \pi R \cos \vartheta \over \beta - \sin \vartheta} \qquad
z = {2 \pi R \sin \vartheta \over \beta - \sin \vartheta}
\ee
Taking the identification into account $z \approx z + 2 \pi R$ and switching to the boosted frame $x'',{\bf y}'',z'',t''$, it is easy to see that on the original brane the effective speed in the boosted frame is
\beq
v = {\Delta x'' \over \Delta t''} = {\gamma \cos \vartheta - B \over 1 - B \gamma \cos \vartheta},
\eeq{mp1}
which is extremized at $\cos\vartheta=\pm1$. At $\cos\vartheta=1$ this formula should coincide with eq.~(\ref{v}) and it does,
in spite of appearances, because of the following elementary computation.
The difference between the two formula for $v$, eqs.~(\ref{v},\ref{mp1}) is
\beq
\gamma \left[{1 + \Gamma^2 B^2 \beta^2 \over 1 - \Gamma^2 B \gamma \beta^2} - {1-B/\gamma  \over 1-B\gamma} \right] = 
\gamma \left[ {(1-B\gamma)(1+\Gamma^2 B^2 \beta^2) - (1-\Gamma^2 B \gamma \beta^2 )(1-B/\gamma) \over (1-\Gamma^2B\gamma \beta^2)(1-B\gamma)}\right] .
\eeq{mp2}
The numerator in~(\ref{mp2}) vanishes identically using the definitions $\Gamma= 1/\sqrt{1-B^2}$, 
$\gamma=1/\sqrt{1-\beta^2}$ since
\bea
\nonumber
&& 1-B\gamma+\Gamma^2 B^2 \beta^2 -\Gamma^2 B^3 \beta^2 \gamma -1 +B/\gamma + \Gamma^2 B\gamma \beta^2-
B^2 \beta^2 \Gamma^2  =   \\
&&= -B\gamma + \Gamma^2 B^2 \beta^2 +B\gamma\beta^2 + B/\gamma -B^2\beta^2\Gamma^2 = -B/\gamma +B/\gamma =0.
\label{mp3}
\eea

Since signalling backwards in time is a slippery concept, let us spell out the observational consequences more fully. 
To that end, imagine that in the far past, long before $t'' = 0$, a grid of brane observers with synchronized clocks has been laid out along the $x''$ axis.  We assume these brane observers use null geodesics on the brane to
synchronize their clocks \cite{Barrow:2001rj}.  These observers define
the $(t'',x'')$ reference frame, which we assume is moving with velocity $B > B_{\rm critical}$ relative to the preferred brane frame. Consider then a signal
emitted at time $t'' = 0$ by a source that is located at $x'' = 0$ and is at rest in the $(t'',x'')$ frame. Much later, observers in the $(t'',x'')$ frame
gather to compare notes.  The observer stationed at the largest positive value of $x''$ will report detecting the signal first, at a time $t'' < 0$ prior to the source at $x'' = 0$
emitting anything.  Observers stationed at successively smaller values of $x''$ will report successively later detections. These are the objective facts that must be explained.

To interpret the data, the observers develop two possible explanations. The first explanation, aligning with the analysis we've presented, is that the source at the origin sent a signal into the past, which caused the cascading series of subsequent detections at ever smaller values of $x''$ at ever later moments in time. The second explanation is that the signal originated at spatial infinity in the infinite past (for reasons unspecified) and traveled in the $-x''$ direction, destined to reach the origin at exactly the moment $t'' = 0$ when the source emitted the signal. 
Either of these two explanations is consistent with the data. The first involves the unfamiliar notion of backward in time signalling. The second, which will appeal to brane observers who view past signalling with suspicion, requires the behavior of the source at the origin--emitting a signal at $t'' = 0$--to be foreordained in the distant past. Either of these interpretations of the experimental data would account for the facts in a surprising but logically consistent manner.  Eventually the observers might develop a third explanation of the data, which is the very scenario we have posited in this paper: that they're living on a brane moving through a higher-dimensional bulk in which signals produced by identifiable sources propagate only forward in time.

We now return to consider the lower part of Fig.\ \ref{fig:cone}. The signal speed arising from the lower part of the envelope (the envelope opposite to the direction we've boosted) is given by
\bea
\nonumber
w & = & {\tan (\theta - \alpha) \over \Gamma \beta} - B \\[3pt]
\label{w}
& = & - \gamma \, {1 + \Gamma^2 B^2 \beta^2 \over 1 + \Gamma^2 B \gamma \beta^2}
\eea
From this we conclude that the effective signal speed is always negative and superluminal.  It varies monotonically from $w = - \gamma$ when $B = 0$ to $w \rightarrow -1$ as $B \rightarrow 1$.  The fact that signals propagate with different
velocities in the forward and backward directions is a clear sign that worldvolume Lorentz symmetry is broken on a moving brane.

It’s worth summarizing how the propagation of the envelope along the brane depends on $B$.  In the sub-critical case $\Gamma^2 B < 1 / \gamma \beta^2$ there is no signal on the brane
for $t'' < 0$.  At $t'' = 0$ a signal appears at the origin and forms an envelope that expands in both the positive and negative $x''$ directions.  In the positive $x''$ direction the envelope moves with velocity $v$
while in the negative $x''$ direction it moves with velocity $w$.  (The signs $v > 0$ and $w < 0$ correctly capture this motion.)
In the super-critical case $\Gamma^2 B > 1 / \gamma \beta^2$  a signal is present on the brane at all times.  It forms an envelope that always travels in the negative $x''$ direction.  The velocity is discontinuous,
changing from $v < 0$ for $t'' < 0$ to $w < 0$ for $t'' > 0$.

\begin{figure}
\centerline{\hspace*{-5mm} \includegraphics[width=8cm]{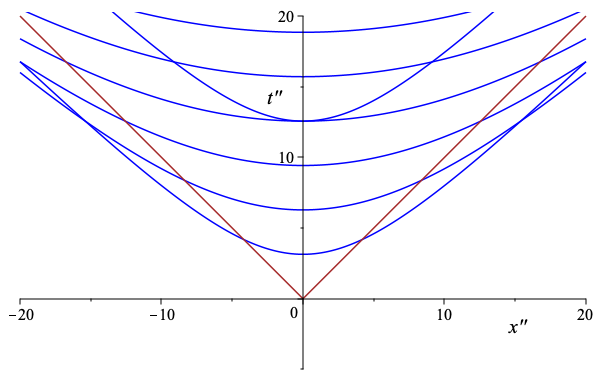} \hspace{6mm} \includegraphics[width=8cm]{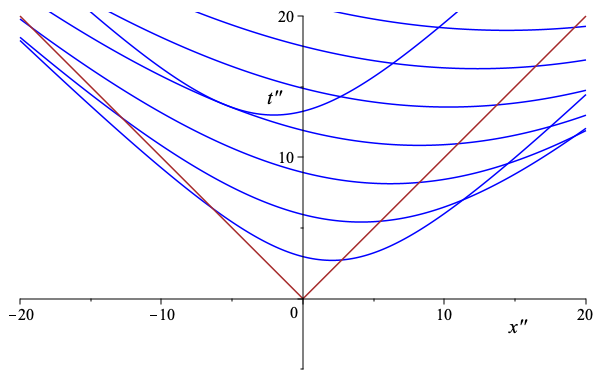}}
\vspace{6mm}
\centerline{\hspace*{-5mm} \includegraphics[width=8cm]{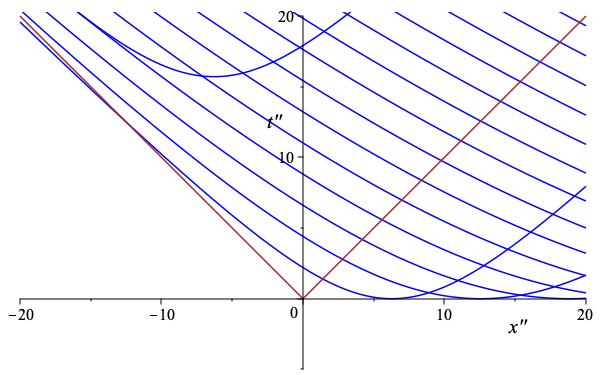} \hspace{6mm} \includegraphics[width=8cm]{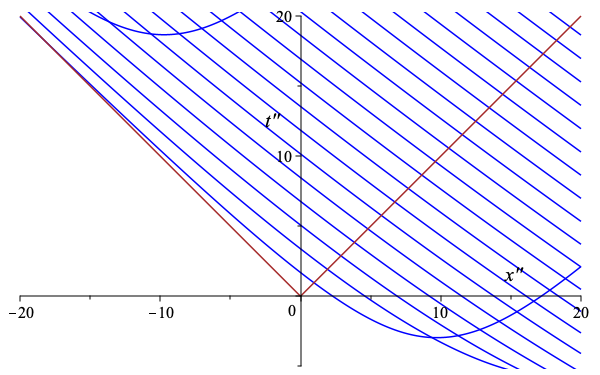}}
\caption{Future lightcones produced by image charges as seen on the brane in the $(x'', t'')$ plane.  The $n = 0$ image charge produces the brane lightcone shown in brown.  Image charges with $n \not= 0$ produce the hyperbolas
shown in blue.  The figures illustrate no boost along the brane $B = 0$, a sub-critical boost $B = 0.4$, a critical boost $B = 0.8$ and a super-critical boost $B = 0.9$.  In all cases $R = 1$ and $\beta = 0.6$.\label{fig:hyperbola}}
\end{figure}

To further illustrate the structure we consider how bulk lightcones produced by image charges appear on the brane.  In the $(x'',t'')$ plane this is given by setting ${\bf y}'' = z'' = 0$ in (\ref{images}).
\be
(x'' - \Gamma B \beta z_n')^2 + (z_n')^2 = (t'' + \Gamma \beta z_n')^2
\ee
The $n = 0$ image charge always produces the lightcone on the brane.  For $\beta > 0$ and $B = 0$ it was pointed out in \cite{Greene:2022urm} that image charges with $n < 0$ produce spacelike hyperbolas nested inside the
brane lightcone while image charges with $n > 0$ (i.e.\ in the direction the brane is moving) produce spacelike hyperbolas that eventually spread outside the brane lightcone.  The effect of the velocity $B$ is simply to boost these
hyperbolas in the $x''$ direction.  Images with $n < 0$ remain within the brane lightcone while images with $n > 0$ can travel into the past.  This is illustrated in Fig.\ \ref{fig:hyperbola}.

\section{Classical causality\label{sect:classical}}
As we noted at the outset, and have now made more apparent, brane motion in a compact direction allows for signals 
to arrive at their destination before they were emitted, flagging a potential issue with causality. A one-way journey by itself cannot violate causality, so to investigate this we need to analyze roundtrip signals.  

Using the envelope velocities (\ref{v}), (\ref{w}), we can calculate the total elapsed time for an out-and-back trip to a point a distance $D$ along the $x''$ axis, given by
\be
T = D \left({1 \over v} - {1 \over w}\right) \equiv {2 D \over v_{\rm eff}}
\ee
where the effective velocity for the round trip is
\be
v_{\rm eff} = 2 \left({1 \over v} - {1 \over w}\right)^{-1} = \gamma \left(1 + \Gamma^2 B^2 \beta^2\right)
\ee
Notice that the effective velocity is superluminal, is bounded below, $v_{\rm eff} \geq \gamma$, and is unbounded above. In particular, $v_{\rm eff}$ and hence $T$ are always positive. Indeed, we have  $0 < T \leq 2 D / \gamma$.  An instantaneous round trip is possible but closed time-like curves are not. In short, there is no causality violation on the brane because the velocity of the return signal, while superluminal, is not superluminal enough.

Causality is a robust feature, certainly more robust than Lorentz invariance which as we've seen is easily violated.  As an instructive additional example, suppose we compactify a spatial direction along the brane by identifying (in the preferred rest frame
on the brane)
\be
\left(\begin{array}{c} t' \\ x' \end{array} \right) \sim \left(\begin{array}{c} t' \\ x' \end{array} \right) +
\left(\begin{array}{c} 0 \\ L' \end{array}\right)
\ee
This turns the spatial direction into a circle.  Does a journey around this circle respect causality?  In a frame that is boosted along the brane as in (\ref{S''}) the identification becomes
\be
\label{circle}
\left(\begin{array}{c} t'' \\ x'' \end{array} \right) \sim \left(\begin{array}{c} t'' \\ x'' \end{array} \right) +
\left(\begin{array}{c} -\Gamma B L' \\ \Gamma L' \end{array}\right)
\ee
The time for a signal to travel around the circle is, making use of (\ref{v}), given by
\be
t'' = {\Gamma L' \over v} = {\Gamma L' \over \gamma} \, {1 - \Gamma^2 B \gamma \beta^2 \over 1 + \Gamma^2 B^2 \beta^2}
\ee
The travel time is negative for $B > B_{\rm critical}$.  However there is a time shift in (\ref{circle}), which when taken into account means the signal returns to $x'' = 0$ after a total time
\be
t'' = \Gamma B L' + {\Gamma L' \over v} = {\Gamma L' \over \gamma} \, {1 + B/\gamma \over 1 + \Gamma^2 B^2 \beta^2}
\ee
This total time is always positive, so like the out-and-back journey considered previously there is no causality violation.  However note that as a function of $B$ the total time is bounded by $0 < t'' \leq {L' \over \gamma}$.
Causality may be safe, but the total travel time can be made arbitrarily small no matter how large the circle is.

\section{Microcausality in QFT\label{sect:microcausality}}
Finally we examine causality, or better microcausality, within quantum field theory.  To this end we consider the expectation value of the commutator
\be
i G(x_1,x_2) = \langle 0 \vert  \, [ \phi(x_1), \phi(x_2) ] \, \vert 0 \rangle
\ee
For a free scalar field in Minkowski space $G$ is related to the retarded Green's function $G_{\rm retarded}$ by \cite{Birrell:1982ix}
\be
G_{\rm retarded}(x_1,x_2) = - \theta(t_1 - t_2) G(x_1,x_2)
\ee
This means we can obtain the commutator simply by suppressing $- \theta(t_1 - t_2)$ in the retarded Green's functions discussed in \cite{Greene:2022urm}.  For a massless scalar field on $M_4 \times S^1$ this
leads to
\be
G(x) \equiv G(x,0) = - {i \over 8 \pi^2} \sum_{n \in {\mathbb Z}} {1 \over \left(\vert {\bf x} \vert^2 + (z - 2 \pi R n)^2 - (t - i \epsilon)^2\right)^{3 / 2}} + {\rm c.c.}
\ee
Here the image sum serves to make $z$ periodic, $z \sim z + 2 \pi R$, and $\epsilon \rightarrow 0^+$ serves to define the singularities in the Green's function.

\begin{figure}
\centerline{\hspace*{-5mm} \includegraphics[height=8cm]{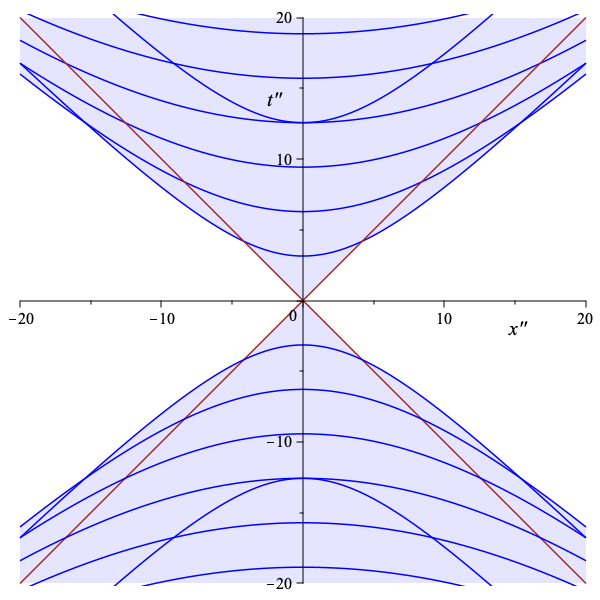} \hspace{6mm} \includegraphics[height=8cm]{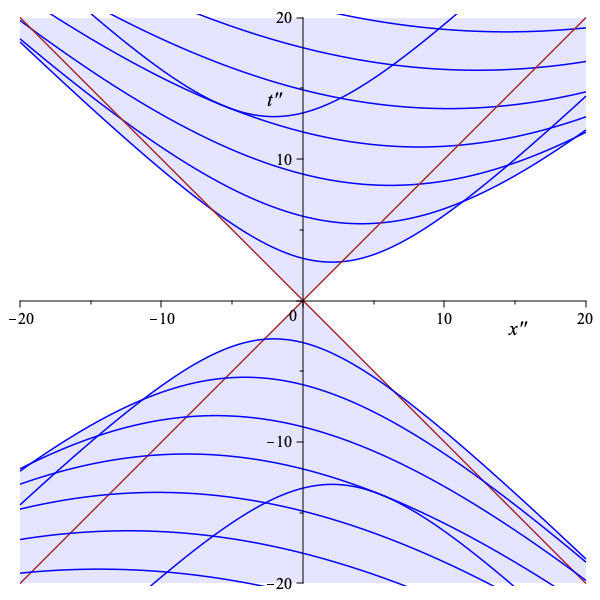}}
\vspace{6mm}
\centerline{\hspace*{-5mm} \includegraphics[height=8cm]{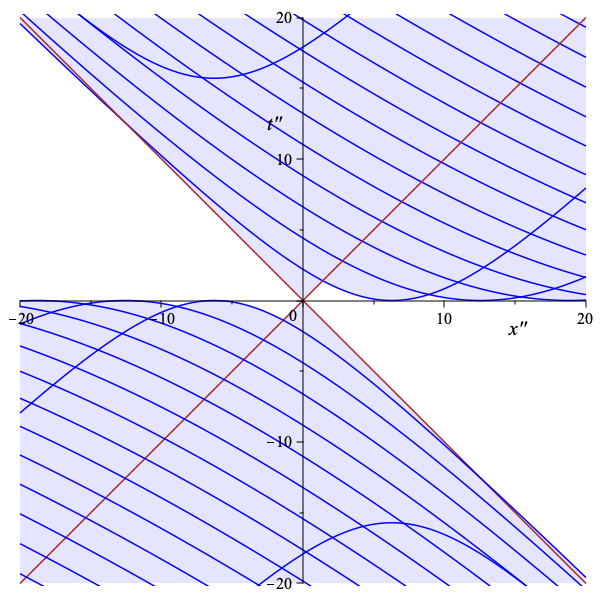} \hspace{6mm} \includegraphics[height=8cm]{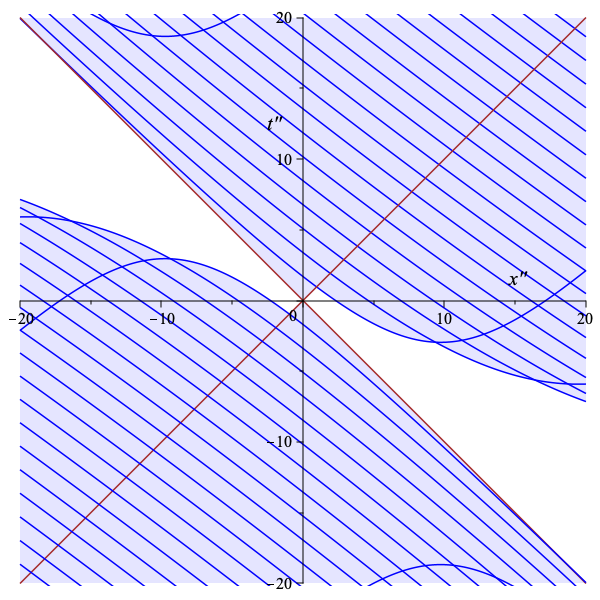}}
\caption{The commutator has support in the shaded regions.  Also shown are the brane lightcone in brown and the upper and lower branches of the hyperbolas for $n \not= 0$ in blue.
Same boosts as in Fig.\ \ref{fig:hyperbola}.\label{fig:commutator}}
\end{figure}

We're interested in evaluating the commutator on the brane, so we switch to a frame $(t',{\bf x}',z')$ that is co-moving with the brane and set $z' = 0$.  This leads to ($z_n' = \gamma 2 \pi R n$)
\be
\label{Gp}
G(t',{\bf x}') = - {i \over 8 \pi^2} \sum_{n \in {\mathbb Z}} {1 \over \left(\vert {\bf x}' \vert^2 + (z'_n)^2 - (t' + \beta z_n')^2 + i \epsilon t'\right)^{3 / 2}} + {\rm c.c.}
\ee
Note that the commutator is non-zero to the future (if $t' > 0$) or past (if $t' < 0$) of the two-sheeted hyperboloids
\be
\vert {\bf x}' \vert^2 + (z'_n)^2 - (t' + \beta z_n')^2 = 0 \qquad n \in {\mathbb Z}
\ee
In these regions the $i \epsilon$ prescription matters and the two terms in (\ref{Gp}) add rather than canceling.  These regions are shaded in the first panel of Fig.\ \ref{fig:commutator}.
We can further boost along the brane, switching to a frame $(t'',x'',{\bf y''})$ that is moving with velocity $B$.  The support of the commutator gets boosted as shown in the
remaining panels of Fig.\ \ref{fig:commutator}.  After boosting the commutator is non-zero to the future of the upper branch of the hyperbolas with $n \geq 0$, a region we will call the causal future.
It's also non-zero to the past of the lower branch of the hyperbolas with $n \leq 0$, a region we will call the causal past.

It's important that the origin has a well-defined causal future and past on the brane.  The two regions meet at the origin but are otherwise separated by an unshaded region.
Lorentz invariance has very much been lost, and the causal future of the origin can include points with $t'' < 0$, but a well-defined notion of causality on the brane survives.
Although we've only considered a free field, we expect causality to be robust with respect to interactions.  Further discussion of this point may be found in \cite{Dubovsky:2007ac}.

\section{Conclusion \label{sect:conclusion}}

We have investigated signal propagation on a braneworld moving along the compact direction of a bulk spacetime $M_4 \times S^1$, from the standpoint of constant velocity observers on the brane itself. Furthering the conclusion of \cite{Greene:2022urm} that such observers will encounter superluminal signal propagation, we have found that--depending on the observer's motion--observers can encounter (i) instantaneous signal propagation or (ii) signal propagation into the past. This raises concerns about causality. But, crucially, we find that all roundtrip signals return to their starting location after a non-negative amount of elapsed time, obviating the possibility of causality-challenging closed timelike curves.

Notwithstanding the gratifying preservation of causality, we find it surprising that simply by compactifying one spatial direction of an otherwise fully Lorentz invariant theory, we enter an arena in which signals can be sent back in time.  Related to this, we can't help noting that interstellar or intergalactic communication in such a universe could in principle take place in real time,  overcoming the usual enormous time delay that would ordinarily thwart such communication in the absence of a compact direction. This is possible even for an observer in the preferred brane frame, $B = 0$, as long as the speed of the brane relative to the preferred frame in the compact direction, $\beta$, is sufficiently close to 1. However, assuming that an observer can only control  their own speed $B$ (and not $\beta$), they still have the capacity to reduce the roundtrip travel time for a light signal, $T$, as much as they desire. For any nonzero $\beta$, as $B$ approaches light speed, $T$ approaches $0$, i.e. the observer can engage in real time communication with an arbitrarily distant partner. 

The direct relevance of this conclusion to our universe depends, of course, on whether we live on a brane, the global nature of the ambient spacetime, the brane's motion through that spacetime, the technological capacity for high speed travel on the brane, and on whether there is anyone out there to hold up their end of such a conversation. Even so, it is surely curious that in such a simple geometrical setting, nearly instantaneous communication across arbitrarily large distances would be possible at all.

\bigskip
\goodbreak
\centerline{\bf Acknowledgements}
\noindent
BG is supported in part by DOE award DE-SC0011941. DK is supported by U.S.\ National Science Foundation grant PHY-2112548 and is grateful to the GGI workshop ``Reconstructing the Gravitational
Hologram with Quantum Information'' for hospitality during this work.  JL is supported in part by the Tow Foundation.
MP is supported in part by NSF grant PHY-2210349 and is grateful to the CERN TH Department for hospitality during the completion of this work.

\providecommand{\href}[2]{#2}\begingroup\raggedright\endgroup

\end{document}